\documentclass[12pt]{article}
\usepackage{epsfig}
\usepackage{amssymb}
\usepackage{amsthm}
\usepackage{amsmath}
\usepackage{amssymb,amsfonts}
\usepackage{graphicx}
\newcommand{\be}{\begin{eqnarray}}
\newcommand{\ee}{\end{eqnarray}}
\renewcommand{\d}{\partial}

\newcommand{\e}{\epsilon}

\newcommand{\Z}{\mathbb{Z}}

\begin{document}
\pagestyle{empty}
\begin{flushright}
UMN-TH-2118/02\\
{\tt hep-th/0211019}
\end{flushright}
\vspace*{5mm}

\begin{center}
{\Large\bf Anomaly-Free Brane Worlds in \\Seven Dimensions}\\
\vspace{1.0cm}

{{\sc Tony Gherghetta}$^a$ and {\sc Alex Kehagias}$^{b,c}$}\\
\vspace{.5cm}
{\it\small {$^{a}$School of Physics and Astronomy,
University of Minnesota,\\
Minneapolis, MN 55455, USA}}\\
{\it\small {$^{b}$ Faculty of Applied Sciences, NTUA, GR-15780
Zografou, Athens, Greece}}\\
{\it\small {$^{c}$ Institute of Nuclear Physics, N.C.R.P.S.
Democritos,\\GR-15310, Athens, Greece}}\\
\vspace{.4cm}
\end{center}

\vspace{1cm}
\begin{abstract}
We present an orbifold compactification of the minimal seven-dimensional
supergravity. The vacuum is a slice of $AdS_7$ where six-branes of
opposite tension are located at the orbifold
fixed points. The cancellation of gauge and gravitational anomalies
restricts the gauge group and matter content on the boundaries. In addition
anomaly cancellation fixes the boundary gauge couplings in
terms of the gravitational constant, and the mass parameter
of the Chern-Simons term.
\end{abstract}

\vfill
\begin{flushleft}
\end{flushleft}
\eject
\pagestyle{empty}
\setcounter{page}{1}
\setcounter{footnote}{0}
\pagestyle{plain}

\section{Introduction}

The geometry of extra spacetime dimensions has recently played
a prominent role in theories beyond the Standard Model. A generic
feature of these models is that one extra dimension is compactified
on a line segment (or orbifold) where boundary worlds exist
at the end points. In particular, the Standard Model gauge and
matter fields are normally assumed to be confined on the boundaries while
gravity propagates in the bulk. It is this geometric separation of the
two boundary worlds, interacting only gravitationally in the bulk, which
has provided new insight into the hierarchy problem~\cite{add,rs}.

In the generic brane world scenario there is no restriction
on the possible gauge group structure on the boundary. Clearly
such a restriction on the gauge group of a higher-dimensional theory
could help to explain the particle content of the low-energy world.
The one notable example is the Horava-Witten (HW) theory
in eleven dimensions~\cite{hw}, where the cancellation of gauge and
gravitational anomalies restricts the gauge group on the ten-dimensional
boundaries to be $E_8$. The bulk eleven-dimensional (11d) theory is then
further interpreted to be the strongly coupled limit of the ten-dimensional
$E_8\times E_8$ heterotic string theory~\cite{hw}.

Apart from ten dimensions, gravitational anomalies also exist in
six (and two) dimensions~\cite{gravanom}. Thus, in this work we
shall show that in seven-dimensional brane worlds the gauge group
structure and matter content is similarly restricted on the 
six-dimensional boundaries by gauge and gravitational anomalies.
Unlike the HW theory where there is a unique gauge group, we will
show that many more possibilities exist in the seven-dimensional
theory, which are not necessarily dimensional reductions of the HW
theory. It should be stressed that, as in the HW case, 
supersymmetry is a central element in our construction. This is because
supersymmetry dictates the possible fields allowed in the bulk as 
well as on the boundaries, and furthermore restricts their possible couplings.

The vacuum of the seven-dimensional theory will be a slice of $AdS_7$, where
six-branes of opposite tension are located at the orbifold fixed points.
This leads to a localized gravity, tensor and hypermultiplet. Moreover,
anomaly cancellation will require the addition of extra vector, tensor
and hypermultiplets on the boundaries. The boundary theory must then
have locally supersymmetric couplings to the seven-dimensional bulk
supergravity multiplet. We will find that the gauge couplings are fixed
by the anomaly cancellation in terms of the bulk gravitational
coupling, and a mass parameter of the bulk Chern-Simons term.

Furthermore, by the AdS/CFT correspondence~\cite{malda},
our seven-dimensional bulk theory is dual to a strongly coupled
six-dimensional (6d) conformal field theory (CFT), in much the same way
that the HW theory is dual to the strongly coupled $E_8\times E_8$
heterotic string theory. This dual correspondence provides a way to further
understand the properties of strongly coupled six-dimensional conformal
field theories.

\section{Seven-dimensional supergravity with boundaries}

Let us consider the minimal ${\cal{N}}=2$ seven-dimensional (7d) gauged
supergravity~\cite{Townsend:1983kk,Mezincescu:ta,Giani:dw}.
The gravity multiplet in this theory consists of the graviton $g_{MN}$,
an antisymmetric three-form $A_{MNK}$, an $SU(2)$ triplet of vectors
${A_M}^{ij}$, a scalar $\phi$, and the $SU(2)$ pseudo-Majorana  gravitino
$\psi_M^i$ and spinor $\chi^i$. A dual version where the three-form is
replaced by a two-form is discussed in~\cite{Salam:1983fa,Han:1985ku}.
The capital Latin indices $M,N=0,1,\dots 6$
are 7d spacetime indices, while $i,j=1,2$ label the $SU(2)$ R-symmetry group.
The bosonic part of the 7d action is
\begin{eqnarray}
&&S_{bosonic}={1\over\kappa^2}\int d^7x \,\sqrt{-g} \, \left[{1\over 2}R
-{\sigma^{-4}\over 48}F_{MNPQ}^2-
{\sigma^{2}\over 4}{{F_{MN}}_i}^j{{F^{MN}}_j}^i-{1\over 2}(\d_M\phi)^2\right.
\nonumber \\
&& ~~~~~ +{i\over 48\sqrt{2}}F_{MNPQ}\left({{F_{KL}}_i}^j{{A_R}_j}^i-{2ig\over
3}{\rm tr}(A_KA_LA_R)\right)\epsilon^{MNPQKLR}\nonumber \\
&& ~~~~~ +\left.60 (m\!-\!{2\over 5}h\sigma^4)^2\!-\!10(m\!+\!{8\over 5}
h\sigma^4)^2
\!+\!{h\over 36} \epsilon^{KLMNPQR}F_{KLMN}A_{PQR}~\right]~,
\end{eqnarray}
where $m=-g \sigma^{-1}/(5\sqrt{2})$, $\sigma=\exp(-\phi/\sqrt{5})$,
and $g$ is the $SU(2)$ gauge coupling.
It can be shown that the  7d supergravity Lagrangian is invariant
under $x_7\to -x_7$ provided that~\cite{gk}
\be
\label{oddtran}
A_{MNP}\to -A_{MNP}\, , ~~~~{A_{Mi}}^j\to -{A_{Mi}}^j\, , ~~~~h\to
-h\, , ~~~~m\to -m~.
\ee
Since the $\mathbb{Z}_2$ transformation, $x_7\to -x_7$ is a symmetry of the
theory, we will consider the compactification down to six dimensions
on the  orbifold $S^1/\mathbb{Z}_2$. Then the only fields which survive
at the orbifold fixed points are the $\mathbb{Z}_2$ singlets, and form the
following 6d multiplets
\be
\label{6dmulti}
&&(g_{\mu\nu},A_{\mu\nu}^+, \psi_\mu^i)\,,
~~~~~~~~~~\mbox{gravity}\nonumber \\&&
(A_{\mu\nu}^-, \phi,\chi^i)\, ,~~~~~~~~~~~~~\mbox{tensor}\nonumber \\&&
(A_i^j,\xi,\psi^i)\, , ~~~~~~~~~~~~~~~\mbox{hypermultiplet} \label{sp6}
\ee
where $A_{\mu\nu}^{\pm}=A_{\mu\nu 7}^\pm$, $A_i^j=A_{7i}^j$, $\xi=g_{77}$,
$\psi^i=\psi_7^i$, and $\psi_\mu^i=\psi_{-\mu}^i$ are left-handed symplectic
Majorana-Weyl fermions while $\chi^i=\chi_+^i$ and $\psi^i=\psi_+^i$ are
right-handed.

The compactification of the 7d supergravity theory on an orbifold results
in a chiral  ${\cal N}=(0,1)$ 6d theory with the massless spectrum
(\ref{6dmulti})
provided that under $x_7\to -x_7$ Eq.(\ref{oddtran}) is satisfied.
However, the relations (\ref{oddtran}) do not necessarily respect the
supersymmetry transformations at the boundaries. For example, since the
parameters $h$ and $m$ are odd, the variation of the kinetic energy terms
will produce $\delta$-function terms. In order to make the truncated theory
on the orbifold supersymmetric
we must introduce six-branes at the orbifold fixed points with
specific boundary potentials. This is very similar to the five-dimensional
supersymmetric Randall-Sundrum model~\cite{gp,susyrs}, where supersymmetry
requires the introduction of brane tensions. If we introduce the boundary
potential term~\cite{gk}
\be
    S_{0} = \int d^6x \int dy \sqrt{-g}\,20(m- \frac{2}{5}h\sigma^4)
        \left[\delta(y)-\delta(y-\pi R)\right]~, \label{Ss}
\ee
then the complete action $S_7 + S_0$ will be supersymmetric, where $S_7$
is the full 7d action including fermionic terms, and $y$ denotes the
seventh coordinate $x_7$.

The supersymmetric vacuum is now the one which satisfies the Killing
equations $\delta\psi_{M i}=\delta\chi_i =0$. Assuming that all bulk fields
are zero except for the scalar field $\phi$, we find that
\be
   \langle \sigma \rangle = \left({g\over 8\sqrt{2} h}\right)^{1\over 5}~.
\ee
In this vacuum the bulk action becomes~\cite{gk}
\be
\label{bbaction}
     S_{bulk}&=&S_7+S_{(0)}+S_{(\pi R)}~,\\
     S_7&=&\int d^6x \int dy \sqrt{-g} \left[ {1\over 2}M^5 R
         - \Lambda_7\right]~,\\
     S_{(y^\ast)}&=& \int d^6x \sqrt{-g_6} \left[ {\cal L}_{(y^\ast)}
        -\Lambda_{(y^\ast)}\right]~,
\ee
where $g_6$ is the induced metric on the six-brane located at $y^\ast$.
The cosmological constants are given by
\be
\label{ftcond}
   \Lambda_7 = -15 M^5 k^2~~ ;~~~~~ \Lambda_{(0)}=-\Lambda_{(\pi R)}
          = 10 M^5 k~,
\ee
where
\be
\label{kdef}
    k= \left({h g^4\over 16}\right)^{1\over 5}~.
\ee
The Einstein equations for the combined bulk and boundary action
(\ref{bbaction}) can be solved to obtain a seven-dimensional
Randall-Sundrum solution
\be
   ds^2 = e^{-2k|y|} dx_6^2 + dy^2~,
\ee
where $0\leq y \leq \pi R$ and $k$ is the AdS curvature scale which
is given by (\ref{kdef}). Note that supersymmetry automatically
guarantees the fine-tuning conditions (\ref{ftcond}) required to obtain
the Randall-Sundrum solution. This leads to a slice of $AdS_7$,
where the 6d gravity multiplet is localized on the UV brane at $y^\ast =0$,
while the tensor and hypermultiplet are localized on the IR brane at
$y^\ast =\pi R$.

\section{Anomaly cancellation with a boundary}

The orbifold compactification of the 7d supergravity theory has resulted
in a theory with a localized gravity multiplet as well as a tensor,
and a hypermultiplet. However, unlike the five-dimensional case where
arbitrary matter can be added to the boundaries~\cite{gp}, in the
slice of $AdS_7$ the 6d fermions of
the vector, tensor and gravity multiplets will in general lead to
gravitational and gauge anomalies. The cancellation of these anomalies
will restrict the possible gauge groups and matter content on the boundary.

In six dimensions the anomalies are formally described by an 8-form, $I_8$.
For $n_V$ vector multiplets, $n_T$ tensor multiplets and $n_H$ hypermultiplets
the requirement for the cancellation of the irreducible tr$R^4$ term in $I_8$,
where $R$ is a curvature 2-form, leads to the condition~\cite{Salam:1985mi}
\be
\label{irredcond}
  n_V-n_H-29 n_T +273 =0~.
\ee
From the dimensional reduction of the bulk gravity multiplet there will
be one tensor multiplet and one hypermultiplet in the 6d theory. However,
this theory by itself is anomalous and we are forced to introduce extra
boundary fields.
In particular we will be interested in introducing vector multiplets on
the boundary which will also produce gauge anomalies.
The gauge group
should be ${\cal G}_1\times {\cal G}_2$, where each ${\cal G}_i$ factor is localized at
the two fixed points (for simplicity we will only consider semisimple
$\cal G$).
Notice that the anomaly need not be equally distributed between the fixed
points as in the HW case. In addition the anomaly eight-form $I_8$
 should satisfy
\be
{\partial^2 I_8 \over \partial {\rm tr}F_1^2\partial {\rm tr}
F_2^2}=0~, \ee
where $F_1,F_2$ are the gauge field strengths of the ${\cal G}_1\times {\cal G}_2$
gauge group. This condition simply means that there is no charged matter
in the bulk and that the only interaction between the two six-branes
is purely gravitational. In this way we will be able to cancel the anomaly
locally on each boundary.

Consider the case of $n_T=1$. Assuming that the
irreducible part of the anomaly tr$R^4$ is cancelled, then the remaining
reducible part  (normalized as in~\cite{Schwarz:1995zw}) is given by
\be
\label{8form}
I_8=({\rm tr} R^2)^2+{1\over 6}{\rm tr}
R^2\left(X_1^{(2)}+X_2^{(2)}\right) -{2\over
3}\left(X_1^{(4)}+X_2^{(4)}\right)~,
\ee
where $X_i^{(n)}={\rm Tr}F_i^n+\sum_in_i{\rm tr}_iF_i^n$, and Tr,
tr$_i$ are traces in the adjoint and the $R_i$ representation, respectively,
whereas $n_i$ is the number of hypermultiplets in the representation $R_i$.
The anomaly (\ref{8form}) can be cancelled by the Green-Schwarz
mechanism~\cite{gs} provided that it can be factorized into the form
\be
\label{global}
   I_8=\left({\rm tr} R^2 + u_i\sum_i {\rm tr} F_i^2\right)
       \left({\rm tr} R^2 + v_i\sum_i {\rm tr} F_i^2\right)~,
\ee
where $u_i,v_i$ are constants. This ensures that at the massless level
the theory is anomaly free.

However, from the 7d orbifold perspective the 6d anomaly must be
distributed between the two fixed points. The bulk topological
Chern-Simons term plays a crucial role in cancelling the anomaly by a
local Green-Schwarz mechanism as occurs in the 11d HW theory.
Thus, by writing $I_8=I_8^{(1)}+I_8^{(2)}$ and demanding the local
factorization
\be
\label{gs}
I_8^{(i)}=\left(c_i{\rm tr}R^2+a_i {\rm tr}F_i^2\right)
\left({\rm tr}R^2+b_i {\rm tr}F_i^2\right)~
\ee
where $a_i,b_i,c_i$ are constants and $c_1+c_2=1$, the two terms in the
sum $I_8$ vanish by a local
Green-Schwarz mechanism at each orbifold fixed point~\cite{Seiberg:1996vs}.
It can be shown~\cite{gk} that the factorization (\ref{gs}) is
indeed possible as long as $\alpha_i {\rm tr}F_i^4=0$.
For $\alpha_i\neq 0$, this occurs for all the
irreps of $E_8,E_7,E_6,F_4,G_2,SU(3),SU(2),U(1)$,
for the {\bf 28} of Sp(4) and SU(8), and all the irreps of SO(2n) with
highest weight $(f_1,f_2,f_1,-f_2,0,...,0)$ in the Gel'fand-Zetlin basis
\cite{Bergshoeff:1986hv}.

Let us now present some examples which illustrate the possible
matter content that is allowed on the six-branes from anomaly cancellation.
In the case where there is only one tensor multiplet in the
6d theory ($n_T=1$), arising from the dimensional reduction of the bulk
theory, we are lead to the constraint
\be
\label{nT1cond}
     n_H=n_V+244~.
\ee
As discussed earlier we will assume that on each boundary there is a
gauge group ${\cal G}_i$. In addition under ${\cal G}_1\times {\cal G}_2$
let us suppose that the total number of hypermultiplets consists of
the following representations
\be
     n_1 (d_{F_1},1) + n_2 (1, d_{F_2}) + (n_S+1) (1,1)~,
\ee
where $d_{F_i}$ is the dimension of the fundamental representation of
the group ${\cal G}_i$, and $n_{1,2},n_S$ are the numbers of each
representation. Note that we have automatically included
the extra singlet hypermultiplet (or radion multiplet) arising from the
dimensionally reduced bulk theory. Thus, assuming the
constraint (\ref{nT1cond}) is satisfied together with (\ref{global}) and
(\ref{gs}), we find the following solutions for ${\cal G}_1={\cal G}_2=
{\cal G}$:
\begin{table}[!h]\centering
\begin{tabular}{|c|c|c|}\hline
${\cal G}\times {\cal G}$ & $n_1+n_2$ & $n_S$ \\ \hline\hline
$G_2\times G_2$ & 20 & 131\\ \hline
$F_4\times F_4$ & 10 & 87\\ \hline
$E_6\times E_6$ & 12 & 75\\ \hline
$E_7\times E_7$ & 8 & 61\\ \hline
\end{tabular}
\end{table}

In particular we see that not only the  gauge groups,
but also the number of hypermultiplet generations is restricted on
the boundaries. For example in the $E_6\times E_6$ case, if one boundary
contains 3 generations of the fundamental ${\bf 27}$ then the other
boundary must have 9 generations. There is also an $(n_1,n_2,n_S)$ solution
$(2,7,156)$, and similar exceptions exist for the other gauge groups.
It is also possible to have two different gauge groups distributed
between the fixed points. These include $E_8\times E_7, E_8\times E_6,
E_8\times F_4, E_8\times G_2, E_7\times E_6, E_7\times F_4, E_7\times G_2,
E_6\times F_4, E_6 \times G_2$, and $F_4\times G_2$.
These exceptions and other possibilities will be presented in Ref.~\cite{gk}.
Our solutions differ from the usual compactifications of the HW theory
because in HW compactifications there is matter charged under
both local gauge groups~\cite{flo,kapetal}. This is true even in
compactifications of weakly coupled string theory~\cite{Green:1984bx}.

Note also that in the $n_T=1$ case, there are no solutions with 
$SU(n), SO(n)$ or $Sp(n)$ gauge groups because the cancellation of the quartic
Casimir does not allow one to simultaneously satisfy
(\ref{global}) and (\ref{gs}). Nevertheless, as will be shown
in~\cite{gk} such solutions can exist for $n_T>1$. However, in
this case one must employ the generalized Green-Schwarz mechanism
of Ref.~\cite{Sagnotti:1992qw}.

Finally note that the six-dimensional theory may still be ill
defined due to non-perturbative anomalies~\cite{Witten:fp,
Elitzur:1984kr,Bergshoeff:1986hv}.  Global anomalies exist as long
as  $\pi_6({\cal G})$ is non-trivial. In our case only the gauge 
group $G_2$ may be plagued by global anomalies since
$\pi_6(G_2)=\Z_3$. In particular, with $n_F$ fundamentals of
$G_2$, the condition for the absence of global anomalies is $n_F=1~~
{\mbox mod}~~ 3$ \cite{Bershadsky:1997sb}. Thus, for the 
$G_2\times G_2$ case, the absence of non-perturbative anomalies further
restricts the values of $(n_1,n_2)$ in the above table.

\section{Bulk-boundary action}

The addition of vector, tensor and hypermultiplets propagating on the
boundaries implies that there must exist locally supersymmetric couplings of
the six-dimensional multiplets to the seven-dimensional supergravity
multiplet propagating in the bulk. In general, the boundary action
can be written in the form $S_{boundary} =  S_0+S_{YM}+S_H+S_T$, where
$S_0$ is given in (\ref{Ss}) and $S_{YM},~S_H,~S_T$ are the boundary
actions for the vector, hyper and  tensor multiplets, respectively.

Let us consider first the addition of vector multiplets on the boundary.
One can show~\cite{gk} that the combined action $S_{bulk}+ S_{boundary}$ is
locally supersymmetric up to fermionic bilinear terms where
\be
S_{YM}\!\!&\!=\!&\!\!-{1\over \lambda^2}\int d^6 x \sqrt{-g}
\left[{\sigma^{-2}\over 4} F_{\mu\nu}^aF^{a\mu\nu}
     \!+{1\over 2}{\bar\lambda}^a\Gamma^\mu
D_\mu\lambda^a+ {\sigma^{-1}\over 4}\bar{\psi}_\mu\Gamma^{\nu\rho}\Gamma^\mu
\lambda^a F_{\nu\rho}^a\right.\nonumber \\
&+& \left.{\sigma^{-1}\over 2\sqrt{5}}\bar{\lambda}^aF^a_{\mu\nu}
\Gamma^{\mu\nu}\chi-{\sigma^{-2}\over 24 \sqrt{2}}\,
\bar{\lambda}^a\Gamma^{\mu\nu\rho}\lambda^a F_{\mu\nu\rho 7}+
{i\sigma\over 2\sqrt{2}}\,\bar{\lambda}^{ai}\Gamma^\mu{F_{\mu 7i}}^j
\lambda^a_j\right]~,
\ee
and  the supersymmetry transformations are
\be
\delta A_\mu^a&=&{1\over 2}\, \sigma\,  \bar{\e}
\Gamma_\mu\lambda^a~,\\
\delta \lambda^a&=&-{1\over 4}\, \sigma^{-1}\, \Gamma^{\mu\nu}
 F_{\mu\nu}^a\e~.
\ee
As in the HW theory one requires the modification of the Bianchi identity for
$F_{\mu\nu\rho7}$. The modified Bianchi identity has the effect
of changing the Chern-Simons term in $S_{bulk}$ into a Green-Schwarz term.
The Green-Schwarz term precisely cancels the anomalous variation of the
effective action for six-dimensional Weyl fermions provided that the
boundary gauge coupling, $\lambda$ satisfies
\be
   \lambda^2= 8\kappa\,  \sqrt{{3\pi^3 h\over \gamma}}~,
\ee
where $\gamma$ is a constant defined by $X_i^{(4)}=\gamma({\rm tr} F^2)^2$.
This relation is similar to that found in the
HW theory, except that now there is an extra dependence on
the topological mass parameter, $h$. In the HW theory the coefficient
of the Chern-Simons term is fixed by supersymmetry whereas in seven
dimensions the theory is supersymmetric up to the arbitrary parameter, $h$.

Similarly we can introduce hypermultiplets on the boundary.
Under the supersymmetry transformations
\be
&&\delta\varphi^\alpha={1\over 2} \sigma^{-1/2}
{V^\alpha}_{iY}\bar{\epsilon}^i\zeta^Y~, \\ &&
\delta\zeta^Y={1\over 2}\sigma^{1/2} {{V_{\alpha i}}^Y}  \Gamma^\mu
\partial_\mu \varphi^\alpha\epsilon^i~,
\ee
the locally supersymmetric boundary action for neutral hypermultiplets
is~\cite{gk}
\be
S_H\!\!\!\! &=&\!\!\!\!\int\!\! d^6x \sqrt{-g}\left[-{1\over 2}\,
g_{\alpha\beta}(\varphi)\partial_\mu\varphi^\alpha\partial^\mu\varphi^\beta
\!-\!{1\over 2}\bar{\zeta}^Y\Gamma^\mu D_\mu\zeta_Y \!
+\!{\sigma^{1/2}\over 2\sqrt{5}} V_{\alpha i Y}\bar{\zeta}^Y
\Gamma^\mu\partial_\mu \varphi^\alpha\chi^i\right.\nonumber\\
&& \left. ~~~~~~~~~~~~~+ \frac{\sigma^{1/2}}{2} \bar{\psi}_\mu^i\
\Gamma^\nu\Gamma^\mu \partial_\nu \varphi^\alpha{V_{\alpha i}}^Y\zeta_Y
+ {\sigma^{1/2}\over  24\sqrt{2}} \bar{\zeta}^Y\Gamma^{\mu\nu\rho}
\zeta_YF_{7\mu\nu\rho}\right]~,
\label{hyperaction}
\ee
where $\varphi^\alpha~(\alpha,\beta=1,...,4n_H)$, and
$\zeta^Y~(X,Y=1,...,2n_H)$ are  the scalars and fermions of the
$n_H$ hypermultiplets, respectively, and $g_{\alpha\beta}$ is the metric
of the scalar manifolds.
Similarly, as in the case of the vector multiplets the Bianchi
identity for $F_{7\mu\nu i j}$ must be modified  which
results in a correction to the supersymmetry transformation of
$F_{\mu 7j}^i$. This is crucial in showing that the scalar manifold
is quaternionic, as required by 6d ${\cal N}=(0,1)$ local
supersymmetry~\cite{Bagger:fn}. Details will be presented elsewhere~\cite{gk}.

\section{Conclusion}

We have presented a new class of models with a boundary, where the
gauge group structure on the boundary is determined by the
cancellation of gauge and gravitational anomalies.
The vacuum of the bulk theory is a slice of $AdS_7$
with localized gravity. Anomaly cancellation also places constraints
on the possible boundary matter, and determines the boundary
gauge coupling in terms of the bulk gravitational constant, and the
mass parameter of the Chern-Simons term.
By the AdS/CFT correspondence our seven-dimensional brane world
is dual to a six-dimensional conformal field theory. Much like the
five-dimensional counterpart~\cite{Ver,Gubser:1999vj,RSholo},
this conformal field theory is defined with a cutoff, and  couples to
gravity. Boundary fields on the UV (IR) brane are identified as
fundamental (composite) states in the CFT, and the strong coupling
regime of this six-dimensional theory is described by our 
seven-dimensional solution.

The six-dimensional boundary theories also have phenomenological interest.
For example, the hierarchy problem is naturally solved and the gauge
group structure on the boundaries can contain the Standard Model gauge group.
Moreover, there are also possible monopole compactifications on $S^2$,
like those considered in~\cite{Salam:1984cj}, which give rise to chiral
four-dimensional ${\cal N}=1$ theories. The low energy particle content
would then be fixed by anomaly cancellation (see also~\cite{6danom}),
and the hierarchy problem is explained by the warped bulk. These issues
as well as cosmological implications remain to be investigated.

\section*{Acknowledgements}
We thank C. Angelantonj, E. Kiritsis, J. March-Russell, E. Poppitz,
R. Rattazzi, A. Sagnotti, J. Sonnenschein, and  R. Sundrum for discussions.
The work of TG was supported in part by a DOE grant DE-FG02-94ER40823 at
the University of Minnesota. The work of AK was supported in part by the
European RTN networks HPRN-CT-2000-00122 and HPRN-CT-2000-00131. TG wishes
to acknowledge the Aspen Center for Physics where part of this work was done.

\end{document}